\documentclass[superscriptaddress,twocolumn,prl]{revtex4-1}

\usepackage{graphicx}
\usepackage{amsmath}
\usepackage{amssymb}
\usepackage{amsthm}
\usepackage[T1]{fontenc}
\usepackage[utf8]{inputenc}
\usepackage{xcolor}
\begin{document}

%---------------------------------------------------------------------------------------------------------
\title{Incipient ferroelectricity: A route towards bulk-terminated SrTiO$_3$.}
%---------------------------------------------------------------------------------------------------------
\author{Igor Sokolović}
\affiliation{Institute of Applied Physics, TU Wien, Wiedner Hauptstrasse 8-10/134, 1040 Vienna, Austria}

\author{Michael Schmid}
\affiliation{Institute of Applied Physics, TU Wien, Wiedner Hauptstrasse 8-10/134, 1040 Vienna, Austria}

\author{Ulrike Diebold}
\affiliation{Institute of Applied Physics, TU Wien, Wiedner Hauptstrasse 8-10/134, 1040 Vienna, Austria}

\author{Martin Setvin}
\email{setvin@iap.tuwien.ac.at}
\affiliation{Institute of Applied Physics, TU Wien, Wiedner Hauptstrasse 8-10/134, 1040 Vienna, Austria}

%---------------------------------------------------------------------------------------------------------
%\begin{abstract} 

%We demonstrate a method for preparing bulk-terminated SrTiO$_3$ (001) by cleaving in vacuum. Strain-induced ferroelectricity plays a key role in the cleaving process. It leads to micrometer-sized domains with SrO and TiO$_2$ terminations that are assigned to  domains of opposite polarization. The as-cleaved surfaces were characterized by combined atomic force/scanning tunneling microscopy (AFM/STM) in the q-Plus configuration and showed (1x1) terminated surfaces wit atomic resolution. Both surface terminations are covered with 14\% of complementary defects that compensate the induced polarity. 

%\end{abstract}

\maketitle

\textbf{Perovskite oxides attract increasing attention due to their broad potential in many applications \cite{Pena2001, Goodenough2004, twdeg1}. 
Understanding their surfaces is challenging, though, because the ternary composition of the bulk allows for multiple stable surface terminations \cite{sr1}.
We demonstrate a simple procedure for preparing the bulk-terminated (001) surface of SrTiO$_3$, a prototypical cubic perovskite.
Controlled application of strain on a SrTiO$_3$ single crystal results in a flat cleavage with micrometer-size domains of SrO and TiO$_2$.
Distribution of these two terminations is dictated by ferroelectric domains induced by strain \cite{ferro4} during the cleavage process. 
Atomically-resolved scanning tunneling microscopy/atomic force microscopy (STM/AFM) measurements reveal the presence of point defects in a well-defined concentration of (14$\pm$2)\%; Sr vacancies form at the SrO termination and complementary Sr adatoms appear at the TiO$_2$ termination.
These intrinsic defects are induced by the interplay between ferroelectricity, surface polarity, and surface charge. 
%The mechanisms demosntrated of SrTiO$_3$ are likely of general validity for other cubic perovskites as well. 
}

%-------------------------------------------------------------------------
\begin{figure*}
    \begin{center}
        \includegraphics[width=2.0\columnwidth,clip=true]{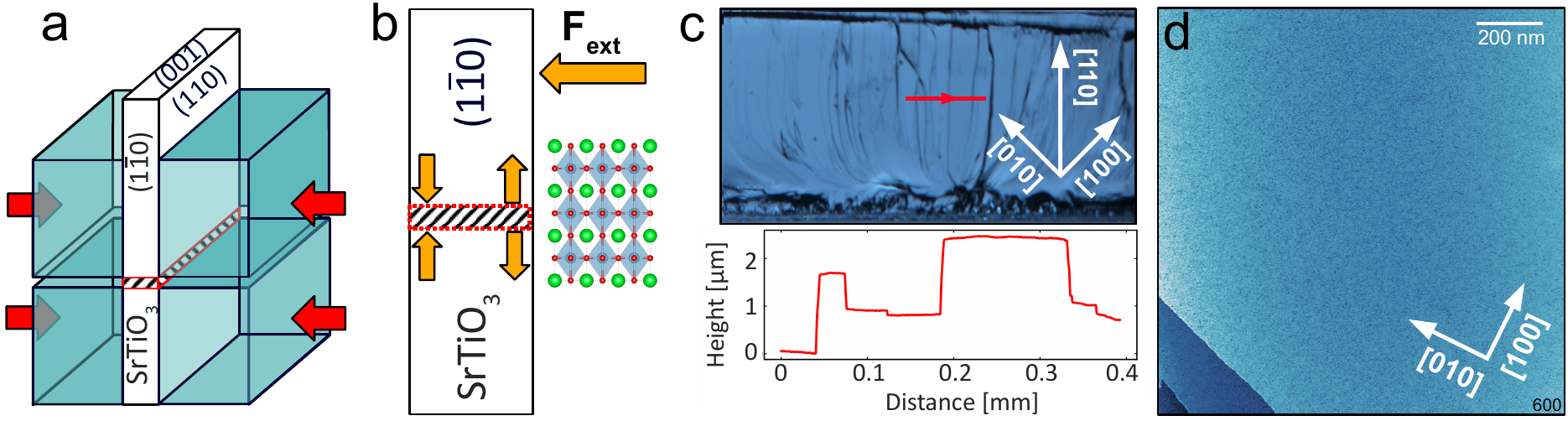}
    \end{center}
\caption{\textbf{Cleaving SrTiO$_3$.} a) Drawing of the sample mount. A SrTiO$_3$ sample is pressed by steel blocks to induce well-defined strain in the cleavage region. b) A sketch of the strain distribution during the cleavage process (orange) and the unit cell distortion (Sr - green, Ti - blue, O - red; the distortion is not to scale) c) Photograph of the as-cleaved surface. A height profile measured along the red arrow shows that all facets have the same (001) orientation. d) Large-scale STM image ($I_\text{T}$ = 2 pA, V$_s$ = 3.3 V,  1.5 $\times$ 1.5 $\mu$m$^2$) of the as-cleaved surface, showing $\mu$m-sized flat regions and steps with a height of one unit cell.}
\end{figure*}
%------------------------------------------------------------------------- 

SrTiO$_3$ is the prototypical cubic perovskite oxide, interesting for its catalytic and photocatalytic properties \cite{cat1}, potential use in oxide electronics \cite{oe1, pg1, pg2, sc1}, and fundamental questions including the appearance of two-dimensional electron gas at its surfaces \cite{twdeg2,twdeg3,twdeg4} and interfaces  \cite{twdeg1}. The SrTiO$_3$(001) surface plays a key role in all these functionalities, but the available surface studies of this facet illustrate a general problem of approaching perovskite materials: A plethora of various surface terminations can form, depending on the preparation technique and on slight changes in the surface stoichiometry. The centre of interest typically lies in bulk-terminated perovskite surfaces, because wet chemical preparation methods typically provide surfaces with a (1$\times$1) diffraction pattern \cite{pg2}. Yet this issue is contoversial due to the possible presence of amorphous structures \cite{Herger2007} or contaminants\cite{pg3}.

Cleaving a single crystal would be an obvious solution for obtaining a pristine surface. Despite numerous attempts \cite{topo4, topo1,topo2}, we are not aware of any successful work leading to atomically flat bulk-terminated SrTiO$_3$.
The main impediment is that cubic perovskites typically do not possess a natural cleaving plane. Instead they exhibit so called conchoidal fracture (see Figure S1 in Supplementary Information). An alternative technique for surface preparation is applying cycles of sputtering and high-temperature annealing - a standard procedure used for preparing oxide surfaces in ultrahigh vacuum (UHV). For perovskite surfaces, however, this methods results into a series of complex reconstructions \cite{sr1, sr4, Kubo2003}, which are very stable and chemically inert; exactly the opposite to the behaviour of perovskites in most applications.

%\section{Results}

%-------------------------------------------------------------------------
\begin{figure}
	\begin{center}
		\includegraphics[width=1.0\columnwidth,clip=true]{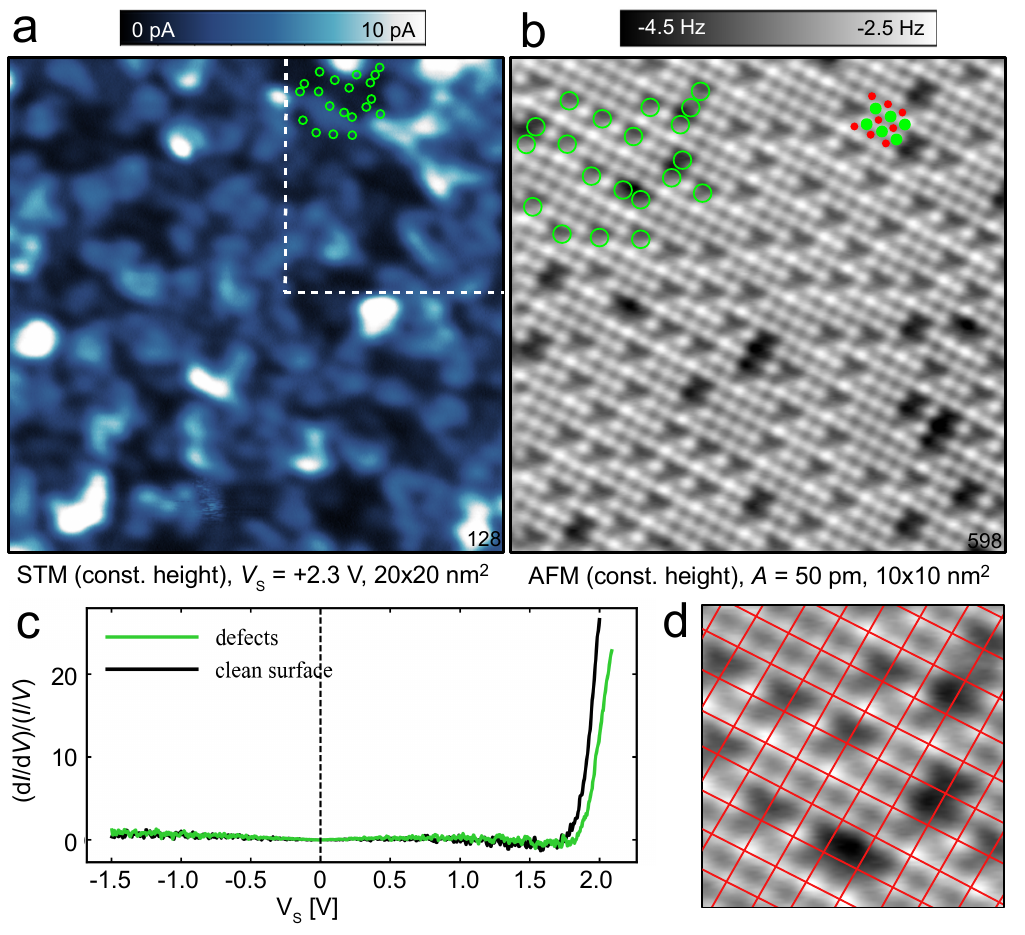}
	\end{center}
	\caption{
		\textbf{SrO termination}
		a) STM image, b) AFM image of the inset marked in (a) with the Sr/O atoms sketched in green/red, and point defects highlighted by the open green circles. c) STS spectra measured above the clean surface and above the point defects. d) Unit cell grid superimposed on the atomically-resolved AFM image.
	}
\end{figure}
%-------------------------------------------------------------------------

%-------------------------------------------------------------------------
\begin{figure}
	\begin{center}
		\includegraphics[width=1.0\columnwidth,clip=true]{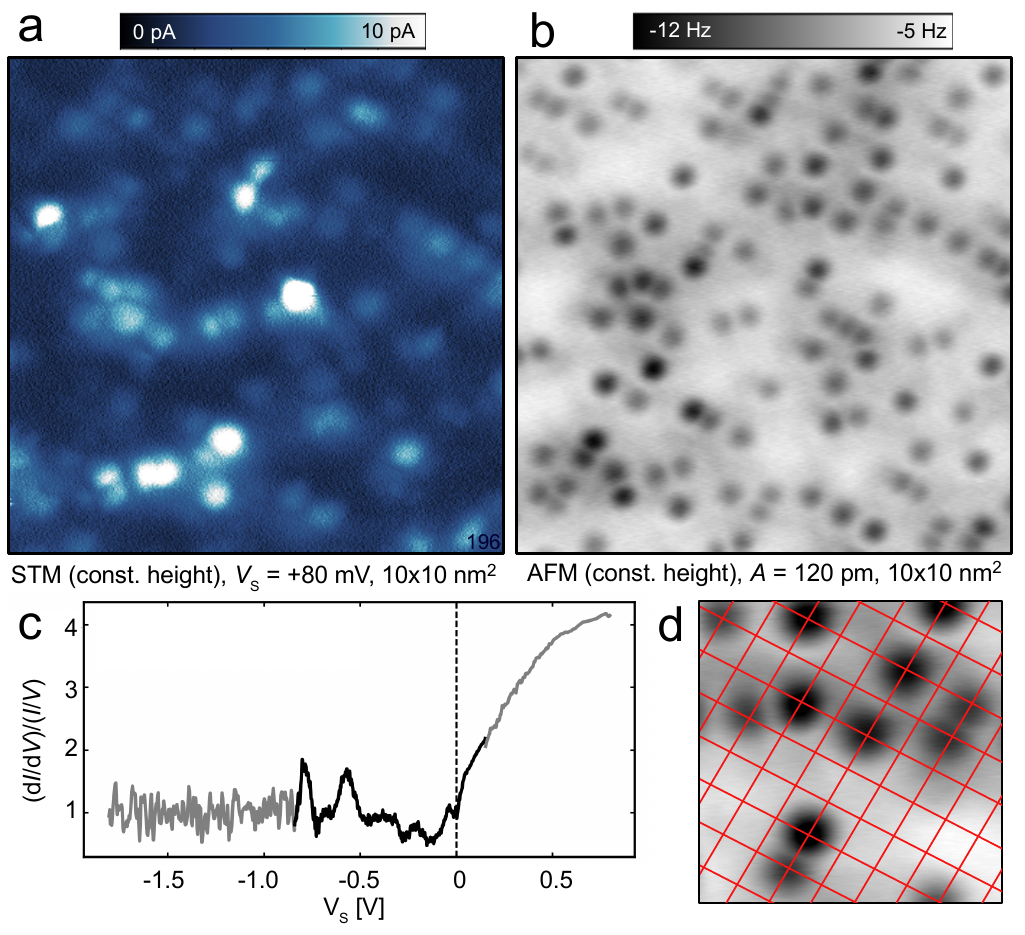}
	\end{center}
	\caption{
		\textbf{TiO$_2$ termination}
		a) STM image, b) AFM image of the same area measured simultaneously. c) STS spectrum of the TiO$_2$ termination. The region close to the Fermi level (black line) was measured at a closer tip-sample distance than the rest of the spectrum (gray). d) Unit cell grid superimposed on the atomically-resolved AFM image.  
	}
\end{figure}
%-------------------------------------------------------------------------

Our previous study on KTaO$_3$ \cite{SetvinScience2018} indicated that ferroelectricity induced in a perovskite can create a natural cleavage plane oriented perpendicular to the electric polarization. SrTiO$_3$ is an incipient ferroelectric material: Its ferroelectric Curie temperature lies below 0~K, but can be shifted even above room temperature \cite{ferro4} when the $c$-axis is sufficiently elongated to suppress the antiferroelectric lattice distortions \cite{Vanderbilt1995,ad1,ad3}. This can be achieved by application of compressive or tensile strain in a direction perpendicular or parallel to the intended polarization direction, respectively. 

Based on these considerations we have designed a device for cleaving the SrTiO$_3$, sketched in Figure 1a. It is based on compressing the STO single crystal along the [110] direction from the side in two regions, while sparing a small fraction of the volume (shaded in Figure 1a) where the cleavage is intended to occur. The cleaving is performed by pushing the crystal ($\mathbf{F}_{\text{ext}}$) along the [110] axis, which induces additional strain, strongly confined in the shaded region (see Figure 1b). In contrast to unstrained crystals, samples properly pre-strained prior to the cleaving become brittle and readily cleave along the (001) plane. A successful cleavage is shown in Figure 1c. The surface is characterized by a train of terraces separated by macrospcopic steps oriented along the [110] direction (parallel to the external strain). A large-area STM image of a SrTiO$_3$ (001) surface cleaved by this method is displayed in Figure 1d. The surface consists of atomically flat terraces ranging from 100 nm up to several $\mu$m in size (see Figure S2 for more examples). 

Using combined STM/AFM, we identified two distinct terminations. Figures 2 and 3 show areas terminated with top layers of SrO and TiO$_2$, respectively. Figure 2a shows a constant-height STM image, the area marked by the dashed square is imaged by AFM in Figure 2b. A clear, bulk-terminated (1x1) pattern is observed in AFM, while the STM image only shows electronic contrast. The lack of atomic resolution in STM is attributed to an absence of localized surface states associated with the atom positions. Superimposed on Figure 2b is a ball model of the (1$\times$1) SrO surface structure. This termination is not perfect as there are point defects in the form of missing atoms. We observe only a single type of vacancies, which, in principle, could be either missing O or Sr atoms. The defect type cannot be determined solely from AFM images, since the contrast achieved on ionic crystals is tip-dependent: Cation-terminated tips interact attractively with surface anions and repulsively with cations \cite{AFM1, AFM2}. An anion-terminated tip will show an inverted contrast (see Figure S3). The unit cell grid superimposed on the atomically-resolved AFM image of the SrO termination (Figure 2d) reveals that all the vacancies appear in equivalent lattice positions, with a concentration of $\eta$(Vac) = (14$\pm$2)\%. This concentration was constant and reproducible in all experiments.

Complementary to the vacancy-riddled SrO termination, the TiO$_2$ terminated surface is covered with point defects in the form of adatoms, seen in Figure 3b. These adatoms are imaged as depressions in the constant-height AFM image, reflecting the attractive chemical force with the tip. The underlying TiO$_2$ lattice is not resolved due to the dominant contribution of the geometrically protruding adatoms. The constant-height STM image (Figure 3a), obtained simultaneously with the atomically-resolved AFM image (Figure 3b), shows that the current signal stems almost exclusively from the adatoms. The brightest species in the STM image (Figure 3a) are probably linked to extrinsic impurities (like Nb dopants). The concentration of the adatoms $\eta$(Ad) = (14$\pm$2)\% equals the previously discussed vacancy concentration on the SrO-terminated surface areas, and the adatoms also sit in equivalent lattice positions (see Fig. 3d).

The electronic properties of the two surface terminations are dramatically different, see the scanning tunneling spectroscopy data in Figures 2c and 3c. The SrO termination appears as a wide-gap semiconductor, with the conduction band minimum located almost 2~eV above the Fermi level. On the other hand, the TiO$_2$ termination is metallic; the Fermi level crosses the bottom of the conduction band and a small density of states is detected throughought the whole band gap. 

Based on their electronic fingerprints, the same concentration, and their specific lattice sites, we attribute the point defects present on both terminations to be strontium vacancies on the SrO termination and complementary strontium adatoms on the TiO$_2$ termination. Sr adatoms act as electron donors and are responsible for the metallic nature of the TiO$_2$ termination. The ionized Sr$^{2+}$ adatoms tend to be distributed across the surface to reduce electrostatic repulsion. Strontium vacancies act as negatively-charged acceptor impurities and are responsible for the upward shift of the SrO conduction band. The locally measured STS spectra above the Sr vacancies exhibit an upward shift of 0.2 V compared to the defect-free regions (the green curve in Fig. 2c). This defect-induced band bending is responsible for the empty-state STM image contrast in Figure 2a: The defects appear as dark depressions (marked by the green circles) because of the upwards band shift. Kelvin probe spectroscopy measurements above the Sr vacancies consistently show a band shift in the same direction, but weaker than in STS (see Figure S4). 

The large-scale morphology of the as-cleaved surface, shown in Figures 4a--c provides valuable insight into the fracturing mechanism behind successful cleavage. The typical domain size of the SrO/TiO$_2$ terminations is much larger than the maximum scan range of our STM/AFM ($>$1~$\mu$m). Interestingly, steps do not switch the termination type and are mostly one unit cell high ($\approx$0.4~nm), or a multiple of that. Figure 4a shows a staircase of atomically flat terraces with TiO$_2$ termination, separated by unit-cell-high steps. Figure 4b shows an area with a more complex morphology containing multisteps; all the terraces here have the SrO termination. The surface termination switches exclusively at very high steps, on the order of 100 nm, which are oriented in the [110] direction (parallel to the applied strain). This is illustrated in Figure 4c, here a multistepstep $\approx$60~nm high separates the SrO termination (left) from the TiO$_2$ termination (right).  

%-------------------------------------------------------------------------
\begin{figure}[t]
	\begin{center}	
		\includegraphics[width=1.0\columnwidth]{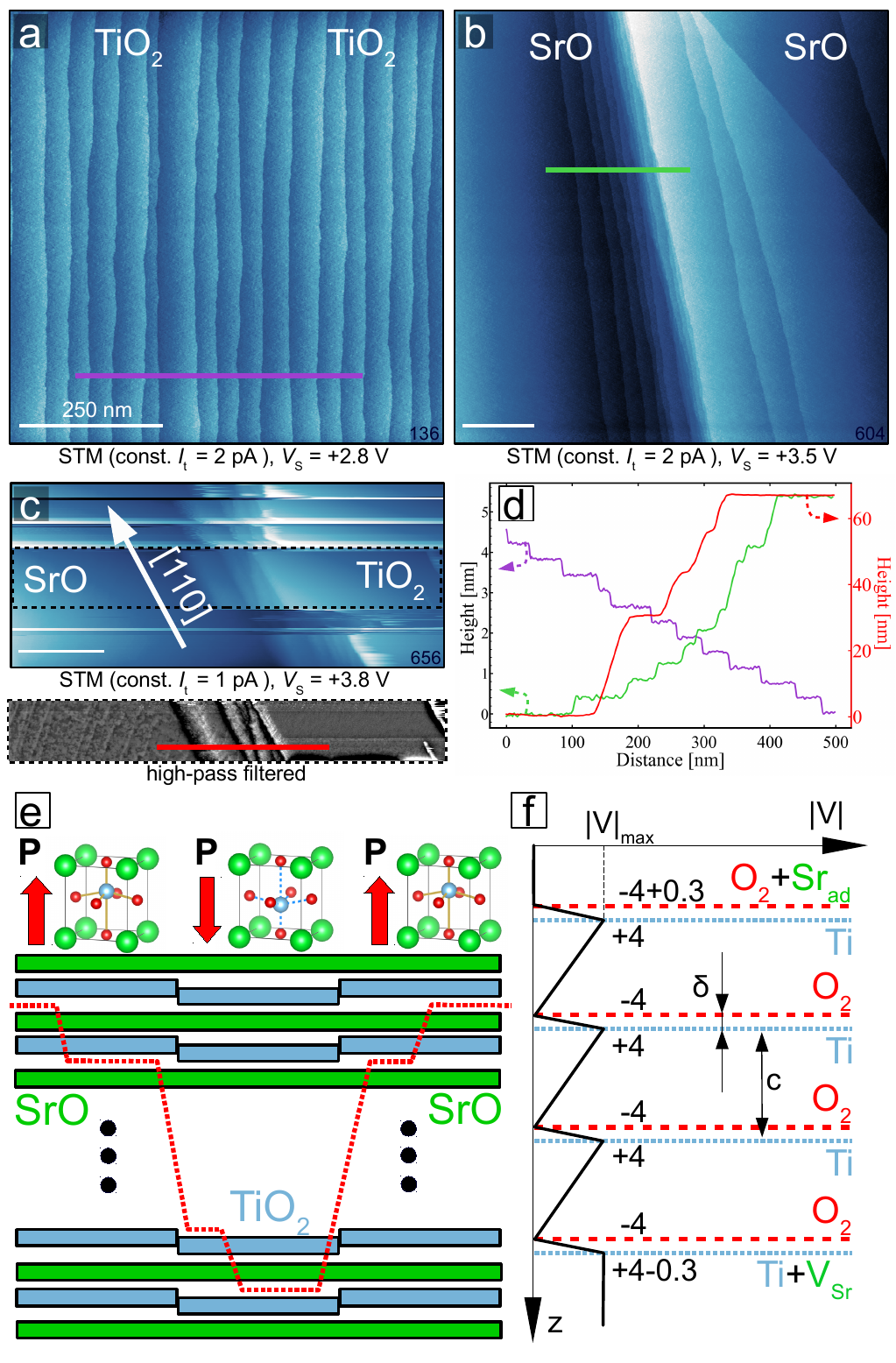}
	\end{center}
	\caption{
		\textbf{Spatial distribution of surface terminations and proposed model for ferroelectricity-assisted cleavage}. a) TiO$_2$ termination with a series of unit-cell-high steps b) SrO termination with a steep 6 nm staircase. c) Change in termination associated with a $\approx$60~nm high step. The region highlighted with a broken line is displayed below after high-pass filtering. d) Line profiles measured along the blue, green, and red lines in panels a,b,c, respectively. e) Proposed scheme of the  induced polarized domains and the fracturing mechanism. The red, dotted line shows the distribution of surface terminations with respect to the polarization orientation. d) Electrostatic potential in a ferroelectric, polar STO crystal with polarization compensation in the form of Sr adatoms and vacancies on opposing surfaces.}
\end{figure}
%-------------------------------------------------------------------------

%\section{Discussion}

The observed distribution of the SrO and TiO$_2$ terminations indicates that large bulk domains might be involved in the cleavage process. We propose a mechanism based on strain-induced ferroelectricity, see Figure 4e. The main feature of the ferroelectric phase in perovskites is an either positive or negative displacement of the B-site transition metal atom along the [001] direction \cite{Vanderbilt1995}. The displacement direction is coherent within a ferroelectric domain and results in a spontaneous electric polarization \textbf{P} along the [001] direction (see the top row of Figure 4e). In a simple electrostatic picture, this distortion separates the formally electroneutral TiO$_2$ layers into two layers with formally equal and opposite charges, Ti$^{4+}$ and (O$_{2})^{4-}$. The distorted TiO$_6$ octahedra create inequivalent (001) planes suitable for cleaving (bottom scheme in Figure 4e), where the resulting surface termination (SrO/TiO$_2$) does not change within the whole ferroelectric domain. The surface termination can only switch at ferroelectric domain walls, which, at the surface, are associated with very large steps. The observed domain size is in the few $\mu$m range, and the orientation of the steps is consistent with observations made on various thin films grown on SrTiO$_3$, where the domain walls run along the main crystallographic directions and linear combinations thereof \cite{dw1,dw3,dw4}.

Our cleaving device is designed to induce the ferroelectric phase transition by a combination of compressive strain from the sample mount, and tensile strain induced by external cleaving force. SrTiO$_3$ requires biaxial compressive strain of 1\% to become ferroelectric at room temperature \cite{ferro4}, or even higher uniaxial compressive strain. In our setup used for $in-situ$ cleaving, the maximum achievable strain is estimated as $\approx$0.4\% (see Figure S5). This is clearly below the value necessary for the ferroelectric phase transition, but it already guarantees a good cleavage. The ferroelectric phase is therefore induced mainly by the high tensile strain associated with the crack propagation during the cleavage \cite{Griffith1921}. The compressive strain from the sample mount therefore serves to confine the tensile strain into a well-defined region, and it also allows for tuning the size of the ferroelectric domains. This is illustrated in Figure S6 on a different perovskite, KTaO$_3$, which forms smaller domains, suitable for STM imaging.   

An interesting feature is the formation of Sr adatoms and vacancies in a well-defined concentration of $\eta$(Sr$_{\text{Ad}}$) = $\eta$(V$_{\text{Sr}}$) = (14$\pm$2)\% in all our successful cleaving experiments. This indicates that they are intrinsic to stabilizing the surface. The ideal STO bulk crystal consists of alternating, weakly polar SrO and TiO$_2$ layers, which would not cause a polar catastrophe when cleaved, since the bond breaking and an increase of bond covalency at the surface layer would serve as a sufficient polarity compensation mechanism \cite{pol1}. However, the cleavage occurs while the crystal is ferroelectric, and here the separation of Ti$^{4+}$ and (O$_2$)$^{4-}$ layers renders the resulting surfaces polar. The removal of Sr$^{2+}$ from the SrO and their addition to the TiO$_2$ termination therefore seems to serve as a mechanism for compenstaing the induced surface polarity \cite{pol2}.  

A simplified model is sketched in Figure 4f. The ferroelectric, polar, STO crystal is represented as stacking of three inequivalent layers along the [001] direction: A negatively charged (O$_2$)$^{4-}$ layer, a positively charged Ti$^{4+}$ layer, and a formally electroneutral SrO layer, which we neglect in the electrostatic considerations. The Sr$^{2+}$ adatoms and V$_\mathrm{Sr}^{2-}$ vacancies are represented by a surface layer with a formal charge of $\pm2\times(0.14\pm0.2) \approx \pm0.3$ electrons per unit cell. The electrostatic potential resulting from the polarization will cancel out with the additional potential from the surface defects: \begin{math} |V|=\frac{4 \delta} {\epsilon}=\frac{0.3 c}{2\epsilon} \end{math}. Here $V$ is the drop of electrostatic potential on one unit cell distance, $\delta$ is the separation of the Ti and O$_2$ layers due to the applied strain, $\epsilon$ is the dielectric permittivity, and $c$ is the lattice constant along the [001] direction. This relation gives an estimate for the distortion $ \delta $ as 3.7\%~$c$, which seems slightly overestimated, but reasonable. The formation of charged surface defects therefore points to an interesting interplay between ferroelectricity and surface polarity. The cleavage occurs when the crystal is in its ferroelectric state, it is likely that the polarization remains in the near-surface region after the cleavage, stabilized by the charge of the surface defects.  

In summary, we have presented a technique for preparing bulk-terminated (001) surfaces of SrTiO$_3$, providing single-domain regions of SrO and TiO$_2$ terminations several micrometers in size. Strain-induced ferroelectricity is the key phenomenon determining the crystal cleavage, and a precise control of the strain allows for controlling the domain size. Results obtained on the KTaO$_3$ crystals (Figure S6) indicate that the technique should be applicable to other incipient-ferroelectric perovskites as well.

%\section{Conclusion}

%In conclusion, cleaving of STO monocrystals was performed by utilizing strain-induced ferroelectricity. Successfully cleaved samples were compressed in such a manner that the  propagation of additional strain during the cleaving is confined in a small region where the cleavage is intended to occur. The exposed (001) surface consists of large, flat terraces covered with a single termination, while a termination change occurs only at the very large steps. The orientation of these large steps indicate that they are the ferroelectric domain walls, mutually separated by several $\mu$m.  The two terminations, SrO and TiO$_2$, have a drastically different electronic structure, making the ferroelectric STO (001) surface a pattern of alternating semiconducting and insulating terraces, respectfully. Both terminations are covered with point defects that compensate polarity, in the form of Sr vacancies on the SrO termination and Sr adatoms on the TiO$_2$ termination. The two surface terminations represent the two ferroelectric domains with opposite orientation of polarization.  The obtained result demonstrates the importance of strain confinement during the cleaving. Moreover, it opens the possibility of successfully cleaving other incipient ferroelectric cubic perovskites and obtaining surfaces normal to the direction of the strain-induced polarization.

%Results

%Discussion

%{ACKNOWLEDGEMENT}

{This work was supported by the Austrian Science Fund (FWF) Projects Wittgenstein Prize (Z 250), Doctoral College "Solids4Fun"   (W 1234), and SFB "FOXSI" (F4505) }

%Experimental details

\section{Materials and Methods}

The AFM/STM measurements were performed in an UHV chamber with a base pressure below 2 $\times$ 10$^{-11}$ mbar, equipped with a ScientaOmicron LTSTM/AFM at T=4.8 K, using the q-Plus setup with a separate wire \cite{Majzik2012} for the tunneling current and a differential cryogenic preamplifier \cite{GiessiblPreamp}. Etched tungsten tips were glued on the tuning fork and cleaned by self-sputtering in Ar atmosphere \cite{SetvinTips2012} prior to the experiment.  The resonance frequency of the used qPlus cantilevers was ~77 kHz with the Q factor of $\approx$ 50000. SrTiO$_3$ samples (CrysTec) with 0.5\% wt Nb doping were used. 

Samples polished along the (110) face were cut into rectangles of 3.5$\times$7 mm$^2$ and 1 mm thickness. The cleavage was performed along the (001) plane, $i.e.$, perpendicular to the polished surface. The profile in Fig. 1c was measured on a Bruker DektakXTL profiler.

%{METHODS DETAILS}

\end{document}